\documentstyle[psfig,aps]{revtex}

\begin{document}

\title{Ordering of atomic mono-layers
on a (001) cubic crystal surface}

\author{ Laurent Proville} 
\address{Groupe de Physique des Solides, UMR 7588-CNRS\\
 Universit\'es Paris 7 $\&$ Paris 6, Tour 23,
 2 pl. Jussieu 75251, Paris Cedex 05, France}

\date{\today}

\maketitle

\begin{abstract}
The self-organization of a chemi-sorbed mono-layer is studied
as a two dimensional ordering process in presence of surface stress.
As proved previously for a single phase separation,
a steady surface state is yielded from the competition between
the domain boundary energy and the surface stress
elastic energy. In the present letter, the 
resulting patterns are shown to depend
on the interplay between the
symmetries of both the internal layer order and the underlying crystal.
For experimental relevance, our study is focussed on a (001) copper surface
and we believe to
enhance a route to stabilize novel surface nanostructures.

\end{abstract}



\twocolumn


The growth of nanostructure onto solid
surfaces provides us with promising technical perspectives
for the electronic miniaturization as for the   
heterogeneous catalyzer assembling. 
The mono-layer self-organization (SO) on crystal surface 
is an efficient mean to control the nanostructure growth
by constructing  a template with
regular nanometer sizes and spacings. The matter which 
may be deposed subsequently on this template 
is likely to organize with same patterns as the mono-layer.

Recent analysis of chemisorbed mono-layers on (001) Copper surfaces, 
via Scanning Tunneling Microscopy (STM) 
\cite{Leibsle,Ellmer,Fishlock}
and Spot Profile Analyzing Low Energy Electron Diffraction (SPA-LEED)
\cite{Croset} showed both a large panel of
morphologies and the means to control their formations.

The interplay between the 
long range elastic interaction yielded by the underlying crystal
surface stress and the domain boundary energy has been well known to control
the surface SO since the papers of 
Marchenko \cite{Marchenko81} and Vanderbilt 
{\it et al.} \cite{Alerland88,Vanderbilt}.
While Refs. \cite{Marchenko81,Alerland88} address the cases of
the vicinal surfaces and the surface reconstruction, 
Ref. \cite{Vanderbilt} was performed in the very general context
of a two-phase system with $1/r^3$
isotropic dipolar interactions in two dimensions 
and thus the latest study is now used to get an insight into
the chemisorbed mono-layer SO. 
Indeed, considering an assembly of surface domains inside which
are fixed adatoms, the energy cost due to the boundary, i.e., where the
adatom environment is unfavorable, is
proportional to the total boundary length L, i.e., I$\times$L. 
As for a given coverage L is minimum for a single compact domain,
the smaller is the number of compact domains, the weaker will be the
domain boundary energy. 
On the other hand, if a non-negligible 
crystal surface stress $\Lambda$ is associated with the adatoms adsorption,
the surface stress inhomogeneities induce some forces
that are located 
at the domain boundaries. These forces yield a crystal strain
and thus an elastic work is involved which 
is minimum when the forces are separated by a
distance as large as possible. So
the surface ground state
structure should balance the twice aforementioned opposite features
and the calculations of 
\cite{Marchenko81,Alerland88,Vanderbilt} proved that periodic domains
occur with a period selection which 
increases exponentially with
the ratio I/$\Lambda^2$, with a suitable multiplier 
which depends on the material elastic constants.

In Ref. \cite{ProSept2001}, a 2-dimensional spinodal theory
was proved to be an efficient tool to study
the SO kinetics on a  (001) cubic crystal surface provided the 
elastic anisotropy due to the underlying crystal symmetries is taken
into account in the calculation of the total free energy $F$.
The 2D Cahn-Hilliard equation was assumed to drive the surface diffusion of the adatoms,
i.e. , the time evolution of the local adatom coverage $\theta$ is given by:

\begin{equation}
\frac{\partial \theta ({\bf r} ,t)}{\partial t}=
M_\theta \bigtriangleup \frac{\delta F }{\delta \theta({\bf r} ,t)} 
\label{KinC}
\end{equation}
A complete analysis of this equation can be found in Refs. 
\cite{Lebowitz,bray} with no elastic interactions, indeed.
The approach developed in Ref. \cite{ProSept2001} 
was actually 
devoted to study a single phase separation on a crystal surface,
no matter how the internal layer
order may play a role. In what follows, we describe how to take into account 
the layer symmetries  of both the adatom layer and 
the underlying cubic crystal. These features
are proved to determine the surface patternings.
We present the different layer 
nanostructures and enhance the control parameters both
for the size and the shape of those structures.
Comparaison with experiments is also proposed as an example of
how to interprete our results.

Some additional order parameters (OP), noted $\eta_j$ are required 
to describe ordered phases that may coexist 
with either orientational or translational variants. 
The kinetics is thus completed by a set
of Allen-Cahn (also known as time dependant Ginzburg-Landau )
equations:
\begin{equation}
\frac{\partial \eta_j ({\bf r} ,t)}{\partial t}=-
M_\eta \frac{\delta F }{\delta \eta_j({\bf r} ,t)} 
\label{KinEta}
\end{equation}
that control the time evolution of  each non-conserved $\eta_j$. 
Such approach was developed
in metallurgical science by Khachaturyan \cite{khacha} 
for the microstructure ordering in alloys. 
The mobility constant $M_\theta$ is  proportional 
to the Fick diffusion coefficient
which is around $10^{-6} cm^2/s$ at 300 K (see \cite{Zangwill}).
As we found no experimental results about the ordering kinetics of surface,
$M_\eta$ is an adjustable parameter which is assumed to fulfill
the adiabatic regime, i.e., the ordering kinetics is 
much faster than the matter diffusion.

The total surface free energy $F$ can be 
written as a sum of two terms, i.e.,
first a chemical term  $F_{chem}$ 
which includes both the energy due to covalent bounds 
between the substrate and the adatoms and the subsequent entropy
and second a
long range elastic term $E_{el}$ due to the crystal surface
stress which is imposed
by the presence of adatoms.
In the framework of a continuous approach, both $F_{chem}$ and $E_{el}$
may be expended with respect
to the coverage $\theta$, the $\eta_j$'s and their respective surface 
gradients. Let first write $F_{chem}$ as follows:

\begin{equation}
F_{chem}=F_0 . \int\int_S \{ \frac{\gamma_\theta}{2}  [\nabla_s \theta]^2 +
\frac{\gamma_{\eta}}{2}  \sum_j [\nabla_s \eta_j]^2
+ {\hat f}(\theta) \}d \mathbf{r}\label{eq8}
\end{equation}
We introduce here the adimensional
free energy density written as
\begin{eqnarray}
\hat f &=& A \theta^2 + E_2 (\theta_1 - \theta) \sum_i \eta_i^2 \nonumber\\
&-& E_{3} \eta_1 \eta_2 -E_4 \sum_i \eta_i^4 + E_5 ( \eta_1 \eta_2)^2
+E_6 \sum_i \eta_i^6  \label{hatF}
\end{eqnarray}
and the surface gradient
$\nabla_s = [(\frac{\partial }{\partial x_1})^2 
+(\frac{\partial }{\partial x_2})^2]$ 
where $(x_1,x_2)$ are the surface coordinates
along the (100) and (010) directions of the (001)
cubic crystal surface.
The  $F_0$  and $\gamma_\eta$, $\gamma_\theta$ scalars are
respectively the free energy density constant and 
the amplitudes of the gradient term that both are adjusted to
set the model domain boundary energy I
to a realistic value, i.e., 
around  $10$ meV/$\AA$ (see Ref. \cite{Alerland88,Prevost}).

\begin{figure}
\noindent
\psfig{file=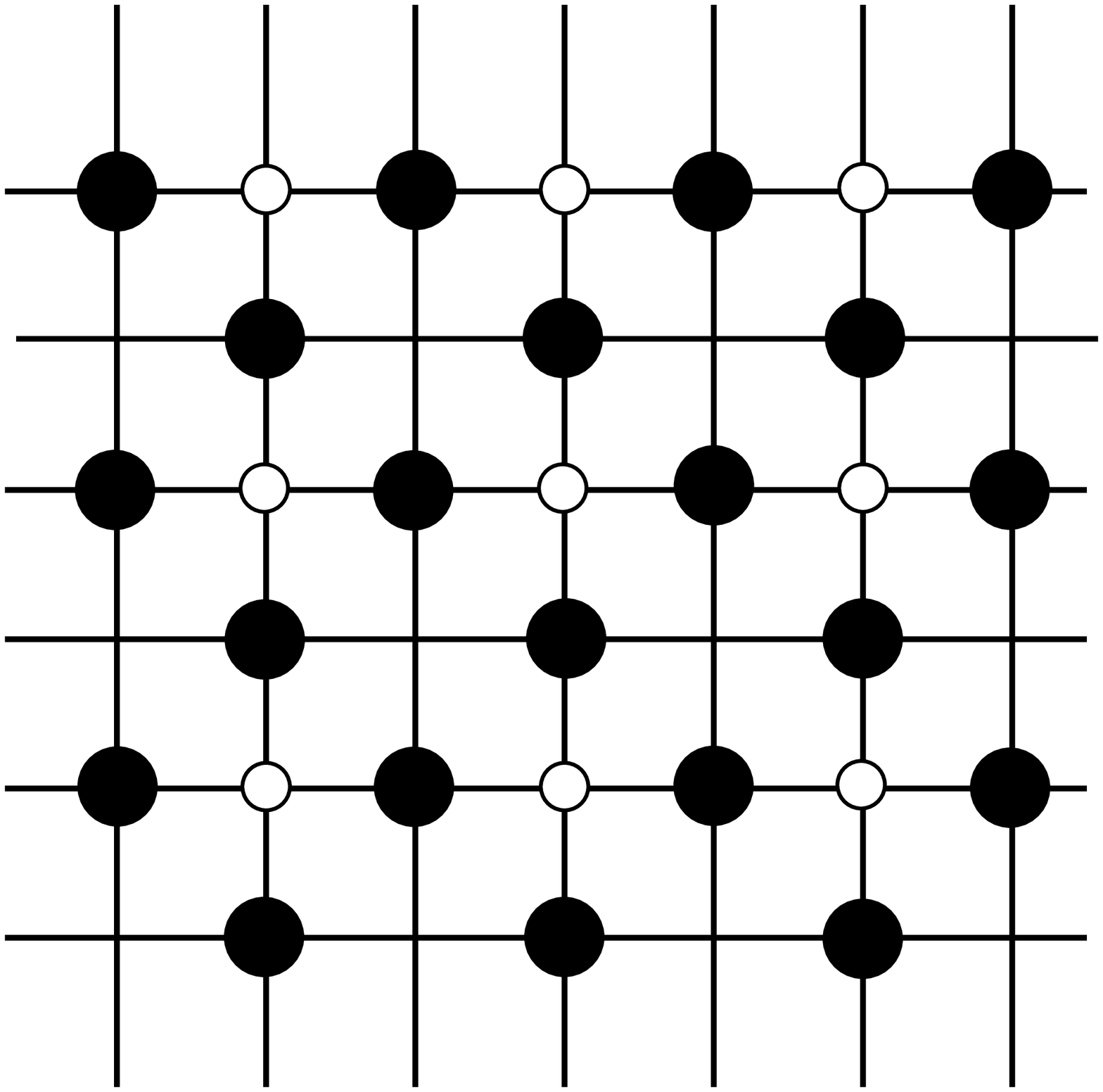,height= 2.5cm,width= 2.5cm}
\psfig{file=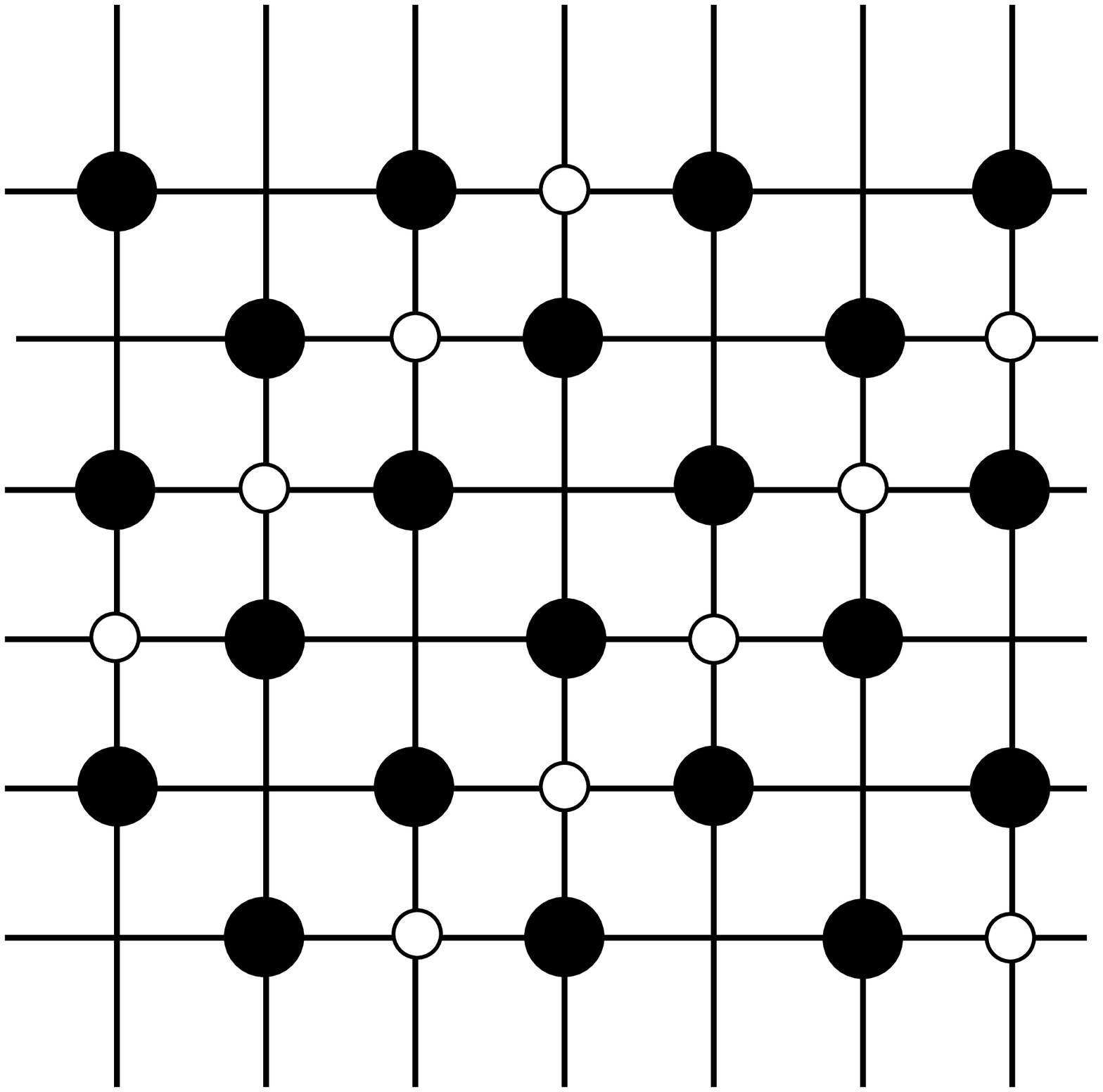,height= 2.5cm,width= 2.5cm}
\caption{\label{schema}: On a (001) fcc crystal surface (atoms of which
are represented by full circles), the adatoms (empty circles)
may arrange in a perfect C2X2 order (on the left hand side) 
or in a P2X1 order (on the right hand side). 
The direction $[0 1 0]$ is indicated.}
\end{figure}

As the OP are supposed to describe the different variants of the
internal layer structure, it is required that 
any symmetry operation relative to this structure should change
the $\eta_j$'s leaving unchanged the $\hat f$ quantity.
For simplicity, the polynomial $\hat f$ expansion is truncated
after the sixth OP power and we focus on a case with only two OP's which is
sufficient to study basic structures such as C2X2 and P2X1, 
well known from surface scientists (see Fig. \ref{schema}
for the case of a (001) fcc surface).

The C2X2 has 2 variants passing from one to the other 
by a [1/2 1/2 0] surface vector translation.
In our formalism, this structure can be represented
by two $\hat f$ minima for $\theta=1$ and with either
$\eta_1=\eta_2= 1$ or $\eta_1=\eta_2= -1$
depending on which variant is considered. 
The internal order P2X1 correspond to 2 orientational variants
since  the adatoms may arrange
either along the direction [110] or [1$\overline 1$0]
and for each orientation there are  two translational variants, 
passing from one to the other 
by a translation of [$1/2$ $1/2$ 0]. 
This structure may correspond to four $\hat f$ minima at
$\theta=1$ and with either $\eta_1=\pm 1$ and $\eta_2=0$
or $\eta_2=\pm 1$ and $\eta_1=0$.
As a result of the P2X1 symmetries, 
the coupling coefficient $E_3$ in Eq. (\ref{hatF}) must be zero.
Minimizing the $\hat f$ potential with respect to OP's
for a given $\theta$ value gives 
two kind of minima, i.e., one disordered 
surface for which the whole set of OP is zero and some 
ordered surfaces for which the OP's have non zero values. 
The $\hat f$ coefficients are adjusted such as
plotting the
$\hat f$ potential after minimizing with respect to the 
OP's gives a double-well
potential with two minima at $\theta =0$ and $\theta=1$.

The $E_{el}$ energy is calculated by inverting the mechanical 
equilibrium equations,
assuming a surface external force distribution $\bf P$. 
At the surface, we have
\begin{equation}
\sigma_{i,j}( {\mathsf{r}},x_3=0). n_j=P_i({\mathsf{r}}) \label{eq1bis}
\end{equation}
where ${ n_j}$ is a  component of the surface normal ${\bf n}=[001]$
and the summation over subscript $j$ is implicit. 
The crystal bulk stress, $\sigma_{i,j}$($\mathsf{ r}$,$x_3$) 
is due to the crystal displacements ${\bf u}({\bf r},x_3)$ 
and it is given by 
the Hooke law: $\sigma_{i,j}=\lambda_{i,j,k,l} \partial u_k / \partial x_l$. 
The forth order tensor
$\lambda_{i,j,k,l} $ gives the crystal elastic constants  and for
a cubic crystal symmetry,
this tensor is composed with three non zero coefficients \cite{Landau}, namely
$\lambda_{i,i,i,i}=C_{11}$, $\lambda_{i,i,j,j}=C_{12}$ and $\lambda_{i,j,i,j}=
\lambda_{i,j,j,i}=C_{44}$.
The bulk displacements fulfill the Lam\'e equation:
\begin{equation}
\lambda_{i,j,k,l} \frac{\partial^2 u_k}{\partial x_j \partial x_l}=0 
\label{eq2}
\end{equation} 
The Eqs.(\ref{eq1bis},\ref{eq2}) are inverted by writing the displacements 
as 2-dimensional Fourier transforms of which
the Fourier components depend on both a surface
wave vector ${\bf Q}=(q_1,q_2)$ and the deepness $x_3$ inside the bulk.
As detailed in \cite{ProSept2001}, it  gives the
surface elastic Green function   $G_{i,l}({\bf Q})$ as a linear function
of the $P_j$ Fourier transform, noted  ${\tilde  P}_j$. 
The total elastic energy of the system is given by
an analytical expression in the Fourier space:
\begin{equation}
E_{el}= - 1/2 \int_{x_3=0} {\tilde P}_i^*  [G_{i,l} ] {\tilde P}_l  d{\bf Q}   \label{eq7}
\end{equation}

Let note $\sigma^0$ the
surface stress imposed by the adsorbed mono-layer. 
The induced force is simply obtained by deriving $\sigma^0$ 
with respect to the surface coordinates which gives:
\begin{equation}
P_i = \sum_{l=1,2} \frac{\partial \sigma^0_{il} }{\partial x_l}
\end{equation}
For simplicity, we choose to 
focus on a case where $P_3=0$ which 
means that $\sigma^0$ can be reduced to a $2\times 2$ matrix.
This stress tensor is expanded with respect to the local coverage $\theta$
and the OP's, i.e.,  $\eta_1$ and $\eta_2$. 
On one hand, if no anisotropy appears in the layer structure
which is the case for a disordered layer
or when there is no orientational variants, e.g., the C2X2,  
then we write
$\sigma^0({\bf r})= \sigma^{00} \theta({\bf r})$ with
$\sigma^{00}_{12}=\sigma^{00}_{12}=0$ and 
$\sigma^{00}_{11}=\sigma^{00}_{22}=\Lambda$.
On the other hand, if there are orientational variants, 
one must add a correction to $\sigma^{00} \theta({\bf r})$
and we propose to write
\begin{equation}
\sigma^0({\bf r})= \sigma^{00} \theta({\bf r})+ \sum_{j=1,2}  \sigma^{0j} \eta_j({\bf r})^2 \label{stressX}
\end{equation}
This expansion holds
when  the OP's correspond one-to-one to the
structure orientations which is the case in our representation of the
P2X1 order. We focus on the P2X1 orientational  
variant where first adatom neighbors are placed along the [110] direction
(Fig. \ref{schema}).
Let note $\Lambda_1$ and $\Lambda_{\overline 1}$, 
the amplitudes  of the stress along [110] and [1${\overline 1}$0], 
respectively. We have $|\Lambda_1 |>|\Lambda_{\overline 1}|$ because
of the proximity of first adatom neighbors which reveals
the internal structure anisotropy.
After performing a suitable  rotation, the stress tensor
is written in the repair  [100]$\times$[010] as follows:
$\sigma^{0}_{11}=\sigma^{0}_{22}=\Lambda$ and
$\sigma^{0}_{12}=\sigma^{0}_{21}=\mu$
where $\Lambda=0.5 (\Lambda_1 + \Lambda_{\overline 1})$ and
$\mu= 0.5 (\Lambda_1 - \Lambda_{\overline 1})$. 
We now identify to the Eq. (\ref{stressX}) for a perfect 
P2X1 ordered domain with the suitable orientational variant
which gives the same expression  
$\sigma^{00}$ as for the C2X2 phase but with
non-zero off-diagonal coefficients for the $\sigma^{01}$ tensor:
$\sigma^{01}_{12} = \sigma^{01}_{21} =\mu$.
Same can be done with the other P2X1 orientational variant and it gives
$\sigma^{02}_{12} =\sigma^{02}_{21} = -\mu$.
As $\Lambda_1$ and  $\Lambda_{\overline 1}$
are assumed to have same sign which means that a dilation (or compression)
occurs in both directions [110] and [1${\overline 1}$0], 
then we note that $|\mu|<|\Lambda|$.

\begin{figure}
\noindent
\psfig{file=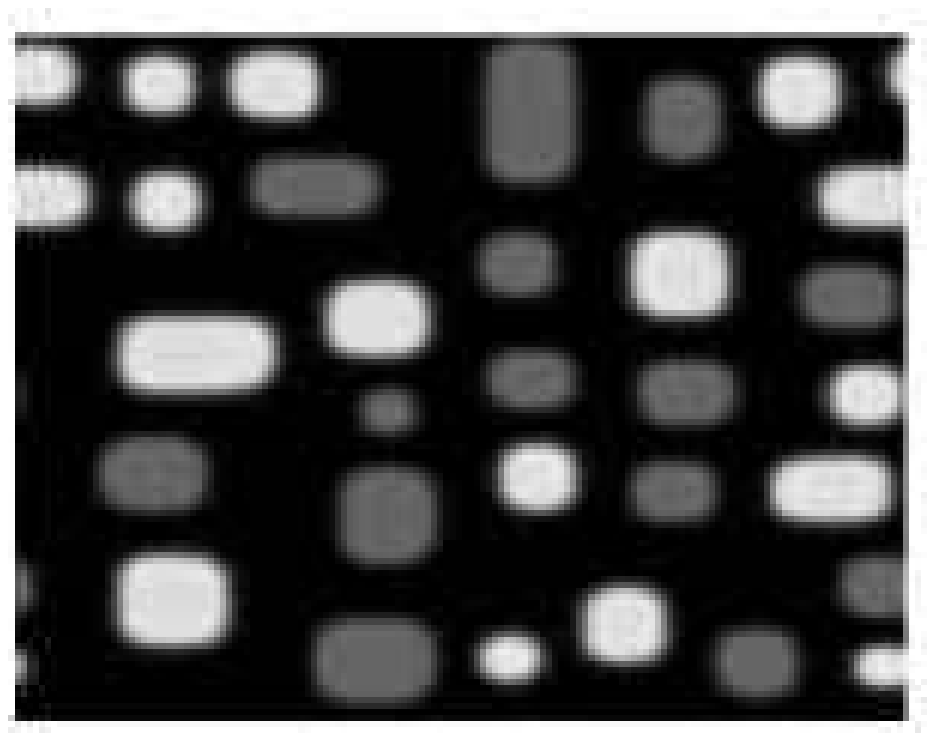,height= 2.7cm,width= 2.7cm}
\psfig{file=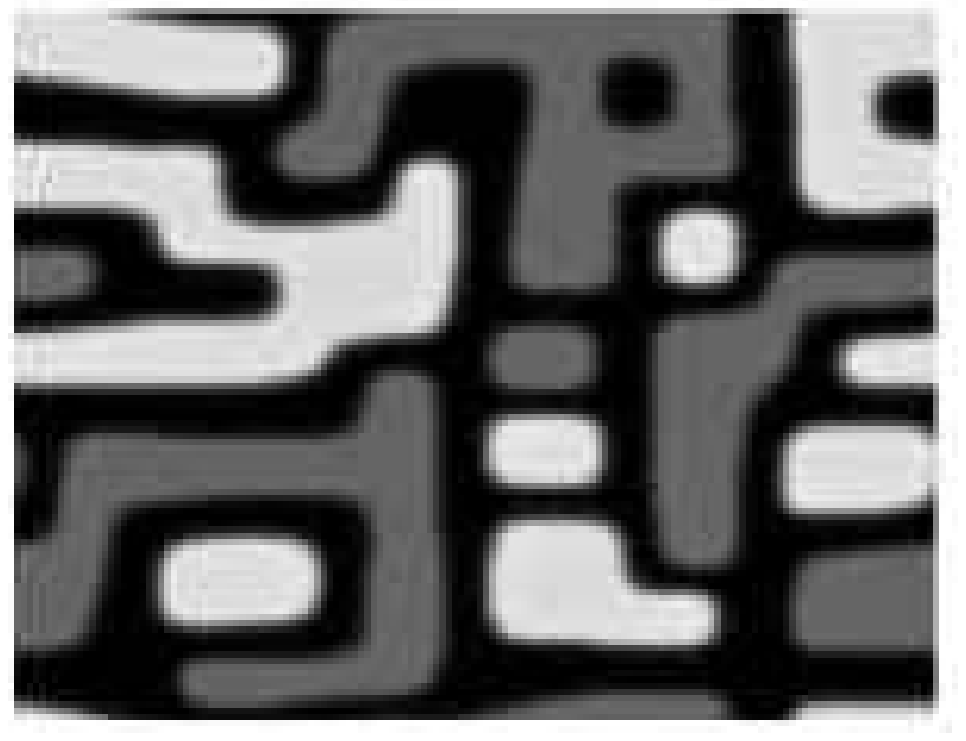,height= 2.7cm,width= 2.7cm}
\psfig{file=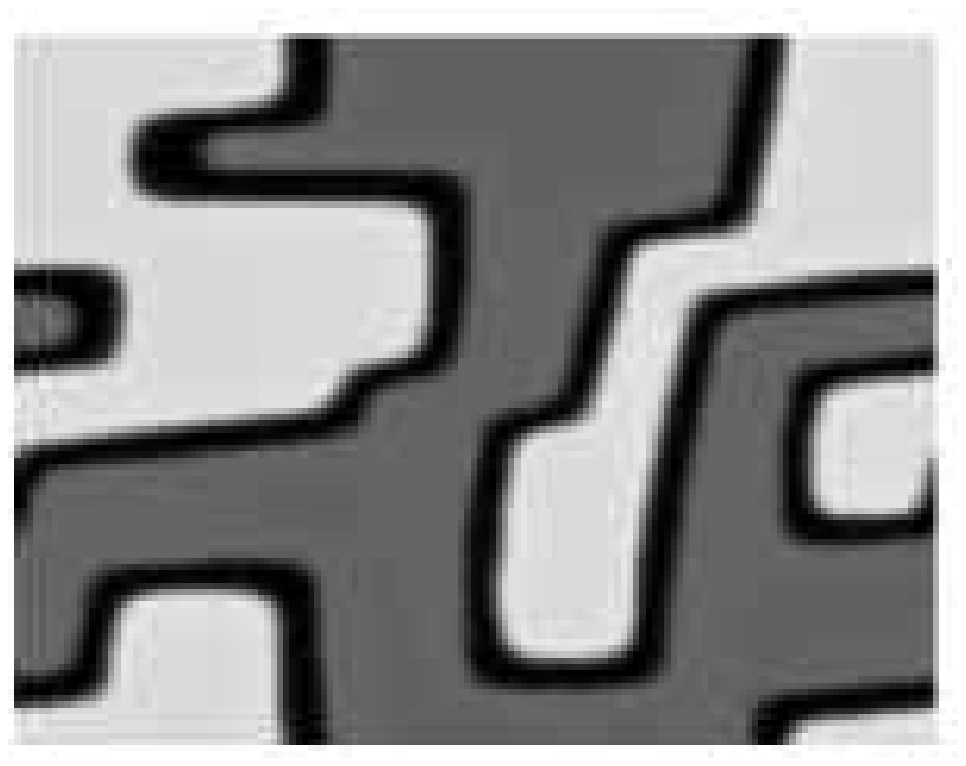,height= 2.7cm,width= 2.7cm}\\
\psfig{file=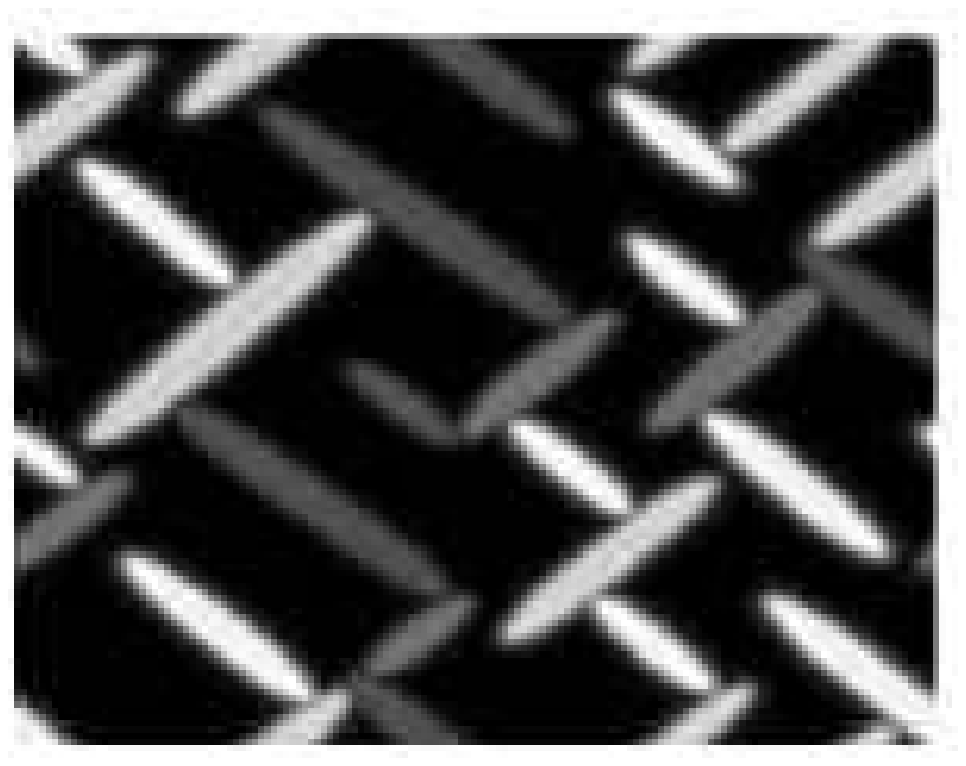,height= 2.7cm,width= 2.7cm}
\psfig{file=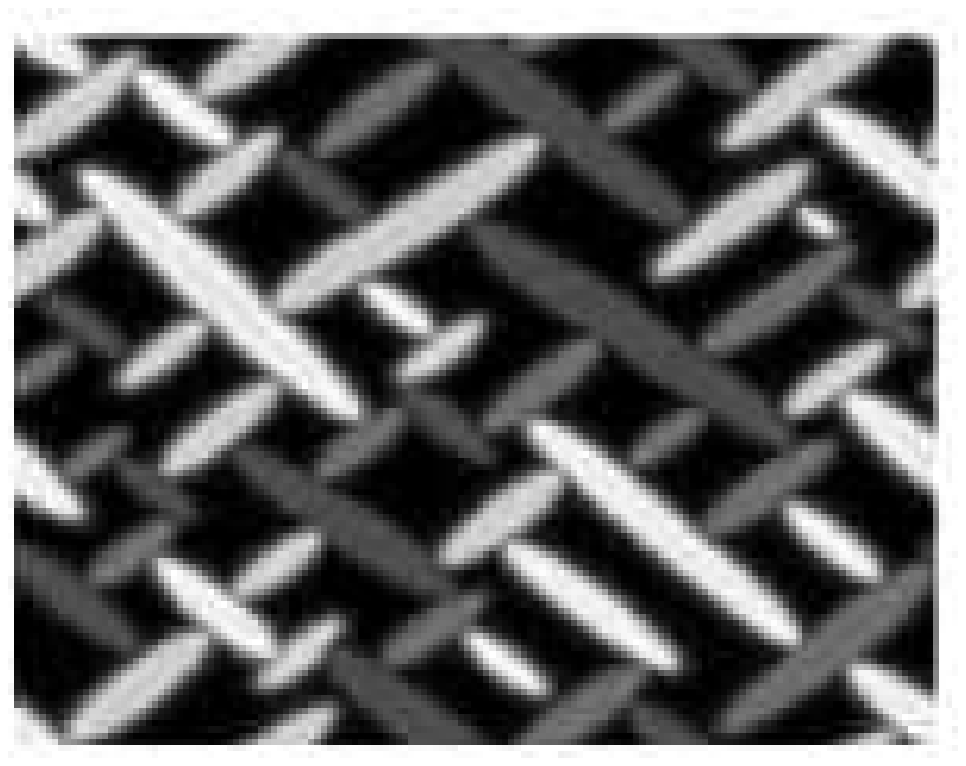,height= 2.7cm,width= 2.7cm}
\psfig{file=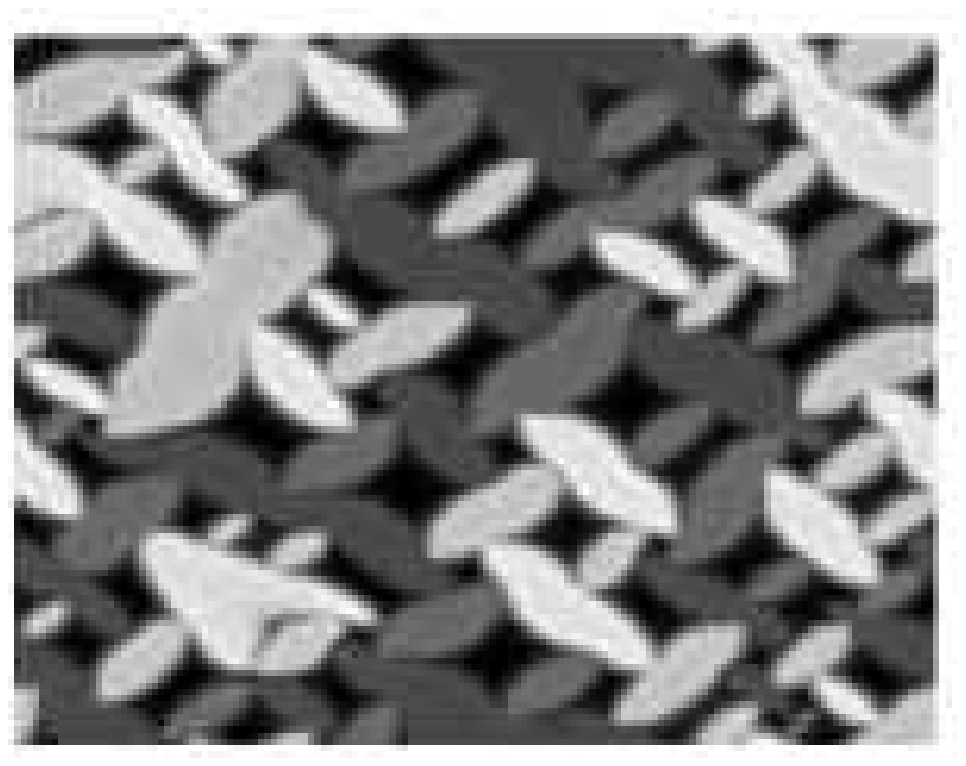,height= 2.7cm,width= 2.7cm}\\
\fbox{\small \bf $[010] \rightarrow $ }
\caption{\label{fig1}: Phase separation final state
on a (001) Copper crystal surface with 
$\Lambda =40\ mJ/m^2 $ (defined in the text) 
for a C2X2   (first row) at 
$\theta_0 = 0.25$ (on the left), $\theta_0 = 0.5$ (in the middle),
$\theta_0=0.75$ (on the right hand side)
and for a P2X1 (second row) with $\mu=0.9\Lambda$ 
at $\theta_0 = 0.38$ (on the left), $\theta_0 = 0.5$ 
(in the middle) and $\theta_0 = 0.75$ (on the right hand side). 
The gray scale enhances the different variants of the layer
structures. 
The direction $[0 1 0]$ is indicated.
}
\end{figure}

Let remind the model parameters:
I=10 meV/$\AA$,  $\Lambda =40\ mJ/m^2 $ ( $\Lambda =0.25 eV/\AA^2$)
and the elastic constants of 
Copper $C_{11}=1.683$, $C_{12}=1.221$ and 
$C_{44}=0.757$ which unit is $10^{11}$ J/m$^3$ (see Physics Handbooks).
The Eqs. (\ref{KinC},\ref{KinEta}) are integrated with a finite 
space and time element method. 
The space unit cell is $1nm \times 1nm$ and the time increment is 
around 1ns.
The kinetics starts from
a uniform coverage $\theta=\theta_0$ and a uniform 
random distribution of OP's between -1 and 1.


On a (001) cubic crystal surface, 
the ordering process
is shown to lead to a steady state 
with different mesoscopic patterns
according to the coverage and the internal structure (see Fig. \ref{fig1}).
Because of the  crystal cubic symmetry and the mono-layer
internal structure, 
the final state differ
from the one predicted in Refs. \cite{Marchenko81,Alerland88,Vanderbilt}.
Nevertheless the space correlation
function of the final state, i.e., $<\theta({\bf r}+{\bf \tau} ) \theta({\bf r})>$ 
exhibits a
characteristic wave length which decreases
exponentially with the boundary energy I which
confirms the predictions
established in \cite{Marchenko81,Alerland88}.

For a perfect C2X2 internal structure, no internal anisotropy 
is induced. Only the symmetries of the 
underlying crystal play a role in the patterning.
At $\theta_0=0.25$, the phase separation kinetics 
yields a nanostructure of square shaped island arranged in raft along
either the [100] or the [010] directions that are the elastic soft
directions of the Copper surface.
The two translational variants are identified with two different shades 
of grey. 
For $\theta_0=0.5$, a
labyrinthine nanostructure occurs with two kinds of wall according to
the  translational variants of the ordered phase. 
For $\theta_0=0.75$, the situation is not the
counterpart of a surface with $\theta_0=0.25$ (see Fig. \ref{fig1})
as one would expect from a simple phase separation. Some
anti-phase boundaries (APB)
due to the coexistence of different order
appear as trenches between the neighboring domains. 
The present theoretical results about 
a (001) Copper surface with a C2X2 mono-layer
may be compared with what is experimentally observed in \cite{Leibsle,Ellmer} 
with the STM analysis of the  N/Cu(001) system. 
The atomic precision of the STM
enhanced adatom missing rows which our model can not capture 
because of the coarse-graining. In the experiments,
those  missing rows occur every 5.2 nm along both [100] and [010] directions
and thus the adatom layer appears as
an assembly of square shaped islands with 5.2 nm size. According to
Leibsle {\it et al.} \cite{Leibsle},  those missing rows are due to
the lattice parameter mismatch between the bulk lattice constants
of Cu$_3$N and the (001) Copper surface unit cell.
Nevertheless if one accepts to
consider the
islands separated by missing rows as a single domain then the 
final state patterns enhanced by the STM experiments 
are very similar to the ones shown in the first row of Fig. (\ref{fig1}), for different coverage. 
Indeed, the C2X2 structure with missing rows
might be consider as a non-perfect C2X2 with no internal anisotropy
as the perfect C2X2.

The P2X1 order implies a layer internal anisotropy, i.e., $\mu\neq 0$. 
In the second row of Fig. (\ref{fig1}),  
it is shown that the patterns strongly differ from the C2X2 case. 
The domains appear as 
tips size of which are oriented along 
specific directions. The weaker is $(1-\mu/\Lambda)$, i.e., the stronger is the internal
anisotropy, the thinner are
the tips and their sizes tends to be align with either [110] or
[1$\overline 1$0].  The second row of Fig. (\ref{fig1}) shows the different patterns
according to the coverage for $\mu = 0.9 \Lambda$ 
which is close from the $\mu$ upper limit.
The tips with different variant
do not branch to each other because of the APB
and the growth of some domains
may be stopped by their neighbors with different orientations.

The anisotropy factor of a cubic crystal
is given by the combination of the elastic constants
$\chi=C_{11}-C_{12}-2C_{44}$ (see Ref. \cite{Landau}). 
The crystal is isotropic for $\chi = 0$
which is the case for amorphous material as glasses.
For example, Copper and Gold $\chi$'s are negative 
$(\chi_{Cu}=-1.0,\ \chi_{Au}=-0.5 )$, and 
Chromium and Niobium $\chi$'s are 
positive $(\chi_{Cr}=+1.8, \ \chi_{Nb}=+0.5)$.
The case $\chi<0$ as for Copper is described above while 
for a positive $\chi$, the soft elastic directions
of the crystal surface  are  [110] and [1${\overline 1}$0] instead of 
[100] and [010] for a negative $\chi$. 
For  $\chi >0$, our calculations showed that the internal anisotropy of a P2X1 layer 
does not modify the preferential orientations of domains
due to the crystal symmetries, i.e.,  [110] and [1${\overline 1}$0].
Only the shape of the domains is changed at low coverage passing from 
square islands arranged in raft along  [110] and [1${\overline 1}$0]
when $\mu=0$ to long tips  aligned in same directions when $ \mu>0 $.
A C2X2 layer deposited on a cubic crystal with $\chi >0$ gives same patterns
as a P2X1 layer with $\mu=0$: the steady surfaces are same as  
the first row pictures  of the Fig. \ref{fig1} after performing a rotation of 45 degrees
around the [001] direction, i.e., 
the domain sizes are aligned with either [110] or [1${\overline 1}$0]
instead of [100] or [010].



In summary, it is proved that the patterning which is yielded by
the ordering of atomic mono-layer onto crystal surface is
controlled by the 
symmetries of both the internal layer structure 
and the underlying crystal. Our results about the 
ordering of a C2X2 layer on a Copper surface are in a good agreement with the experiments.
The method we used in the present paper is a powerful tool to predict
the nanostructures that may be constructed experimentally with different
crystal and different adsorbate. 

In an other field of research, i.e., the magnetic films,
a similar approach have been used by A. Cebers \cite{Poland} who enhanced the different  patterns
that may be yielded from the interplay between the magnetic domain boundary energy and the long range magnetic 
dipolar energy, according to the external magnetic field. To that respect,
we believe that a  model extension including both the magnetic
and the elastic energy terms should be possible
to describe the magneto-striction which may occur on a crystal surface
by deposing magnetic compounds.




\end{document}